\documentclass[fleqn,12pt,twoside]{article}
\usepackage{espcrc1}

\usepackage{graphicx}

\begin{document}
\title{Event-by-event fluctuations at SPS}

\author{
Harald Appelsh\"{a}user and Hiro Sako for the CERES
Collaboration:\\
~\\
D.~Adamov\'a$^{\rm a}$,
G.~Agakichiev$^{\rm b}$,
H.~Appelsh\"auser$^{\rm b}$,
V.~Belaga$^{\rm c}$,
P.~Braun-Munzinger$^{\rm b}$,
A.~Castillo$^{\rm b}$,
A.~Cherlin$^{\rm d}$,
S.~Damjanovi\'c$^{\rm e}$,
T.~Dietel$^{\rm e}$,
L.~Dietrich$^{\rm e}$,
A.~Drees$^{\rm f}$,
S.\,I.~Esumi$^{\rm e}$,
K.~Filimonov$^{\rm e}$,
K.~Fomenko$^{\rm c}$,
Z.~Fraenkel$^{\rm d}$,
C.~Garabatos$^{\rm b}$,
P.~Gl\"assel$^{\rm e}$,
G.~Hering$^{\rm b}$,
J.~Holeczek$^{\rm b}$,
V.~Kushpil$^{\rm a}$,
B.~Lenkeit$^{\rm g}$,
W.~Ludolphs$^{\rm e}$,
A.~Maas$^{\rm b}$,
A.~Mar\'{\i}n$^{\rm b}$,
J.~Milo\v{s}evi\'c$^{\rm e}$,
A.~Milov$^{\rm d}$,
D.~Mi\'skowiec$^{\rm b}$,
Yu.~Panebrattsev$^{\rm c}$,
O.~Petchenova$^{\rm c}$,
V.~Petr\'a\v{c}ek$^{\rm e}$,
A.~Pfeiffer$^{\rm g}$,
J.~Rak$^{\rm b}$,
I.~Ravinovich$^{\rm d}$,
P.~Rehak$^{\rm h}$,
H.~Sako$^{\rm b}$,
W.~Schmitz$^{\rm e}$,
J.~Schukraft$^{\rm g}$,
S.~Sedykh$^{\rm b}$,
S.~Shimansky$^{\rm c}$,
J.~Sl\'{\i}vov\'a$^{\rm e}$,
J.~Stachel$^{\rm e}$,
M.~\v{S}umbera$^{\rm a}$,
H.~Tilsner$^{\rm e}$,
I.~Tserruya$^{\rm d}$,
J.\,P.~Wessels$^{\rm i}$,
T.~Wienold$^{\rm e}$,
B.~Windelband$^{\rm e}$,
J.\,P.~Wurm$^{\rm j}$,
W.~Xie$^{\rm a}$,
S.~Yurevich$^{\rm a}$,
V.~Yurevich$^{\rm a}$,
}


\maketitle
~\\
$^{\rm a}$Nuclear Physics Institute ASCR, 25068 \v{R}e\v{z}, Czech Republic\\
$^{\rm b}$Gesellschaft~f\"{u}r~Schwerionenforschung~(GSI),~64291~Darmstadt,~Germany\\
$^{\rm c}$Joint Institute for Nuclear Research, 141980 Dubna, Russia\\
$^{\rm d}$Weizmann Institute, Rehovot 76100, Israel\\
$^{\rm e}$Physikalisches Institut der Universit\"{a}t Heidelberg, 69120
Heidelberg, Germany\\
$^{\rm f}$State University of
New York--Stony Brook, Stony Brook, New York 11794-3800\\
$^{\rm g}$CERN, 1211 Geneva 23, Switzerland\\
$^{\rm h}$Brookhaven National Laboratory, Upton, New York 11973-5000\\
$^{\rm i}$Institut f\"{u}r Kernphysik der Universit\"{a}t
M\"{u}nster, 48149 M\"{u}nster, Germany\\
$^{\rm j}$Max-Planck-Institut f\"{u}r Kernphysik, 69117 Heidelberg, Germany\\

\begin{abstract}
Results on event-by-event fluctuations
of the mean transverse momentum and net charge in Pb-Au 
collisions, measured by the CERES Collaboration at CERN-SPS,
are presented. We discuss the centrality and beam energy
dependence and compare our data to cascade calculations.

\end{abstract}

\section{\label{sec:level1} Introduction}

In collisions of heavy nuclei at high energies, the creation
of hot and dense matter consisting of deconfined quarks and gluons 
is expected.
If thermalization is achieved in the early stage of the collision,
a phase transition to a Quark-Gluon-Plasma (QGP) may occur, as indicated
by QCD calculations on the lattice.
The experimental verification of the phase transition based on the
final state distribution of the produced hadrons is, however,
rather circumstantial. 
Non-statistical fluctuations of
intensive quantities, such as the event-by-event 
mean transverse momentum $M_{pt}$
of hadrons, have been proposed as a possible signature for 
critical phenomena connected with the passage of the system
through the phase boundary, and may therefore serve as an important 
tool for the exploration of the QCD phase diagram~\cite{steph1,dum}.
The formation of an equilibrated QGP may also lead to reduced fluctuations
of the net electric charge in the hadronic final state~\cite{asa,jeo}. 
This expectation
arises from the smaller charge units of quarks compared to pions 
and the large contribution
to charged particle production by gluons, which are electrically neutral
and do not contribute to fluctuations.

In this contribution, results on event-by-event fluctuations 
are presented from Pb-Au collisions at 40, 80,
and 158 $A$GeV. The data are recorded
with the TPC of the CERES experiment at
CERN-SPS.
The TPC covers the polar angle range $8^{\circ}<\theta<15^{\circ}$,
corresponding to about half a unit in pseudorapidity close to mid-rapidity.

\subsection{Fluctuations of the mean transverse momentum}

Fluctuations of the event-by-event mean transverse momentum
$M_{pt}$ are composed of statistical fluctuations arising
from the finite number of tracks per event, and a possible non-statistical
({\em dynamical}) contribution.
The mean transverse momentum $M_{pt,j}$ of event $j$ with $N_j$
charged particles is defined as 
$	M_{pt,j} = \frac{\sum_{i=1}^{N_j}p_{t,i}}{N_j}$.
As a measure for event-by-event fluctuations, we 
employ the quantity 
$\Sigma_{pt}$~\cite{voloshin,ceresnpa} which is dimensionless and specifies the 
dynamical contribution to event-by-event $M_{pt}$ fluctuations
in fractions of the inclusive mean transverse momentum 
$\overline{p_t}$. In the case of independent
particle emission from a single parent distribution, $\Sigma_{pt}$ is zero.

The finite two-track separation of the TPC leads to a suppression
of particle pairs with small momentum difference and consequently
to a slight anti-correlation of particles in momentum space.
In the case of the CERES-TPC, the effect on $\Sigma_{pt}$
is negligible~\cite{ceresnpa},
hence no correction has been applied. Positive correlations
may arise due to quantum statistics, flow, jets, and other {\em physics} 
effects which have also not been corrected for.

\begin{figure}[htb]
\begin{minipage}[t]{84mm}
\includegraphics[width=\textwidth]{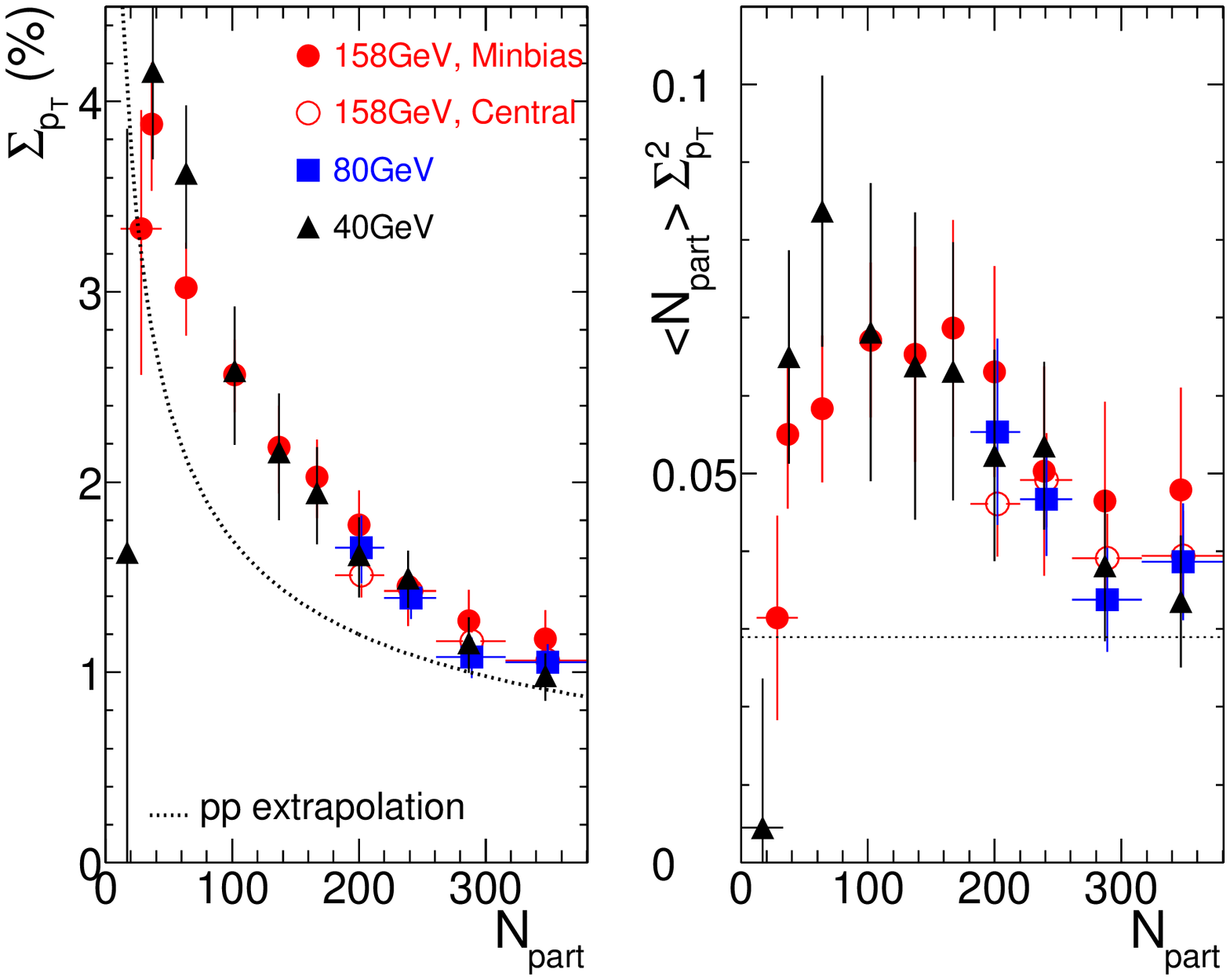}
\caption{Centrality dependence of $\Sigma_{pt}$ at 40, 80, and 158~$A$GeV/$c$.}
\label{fig1}
\end{minipage}
\hspace{\fill}
\begin{minipage}[t]{75mm}
\includegraphics[width=\textwidth]{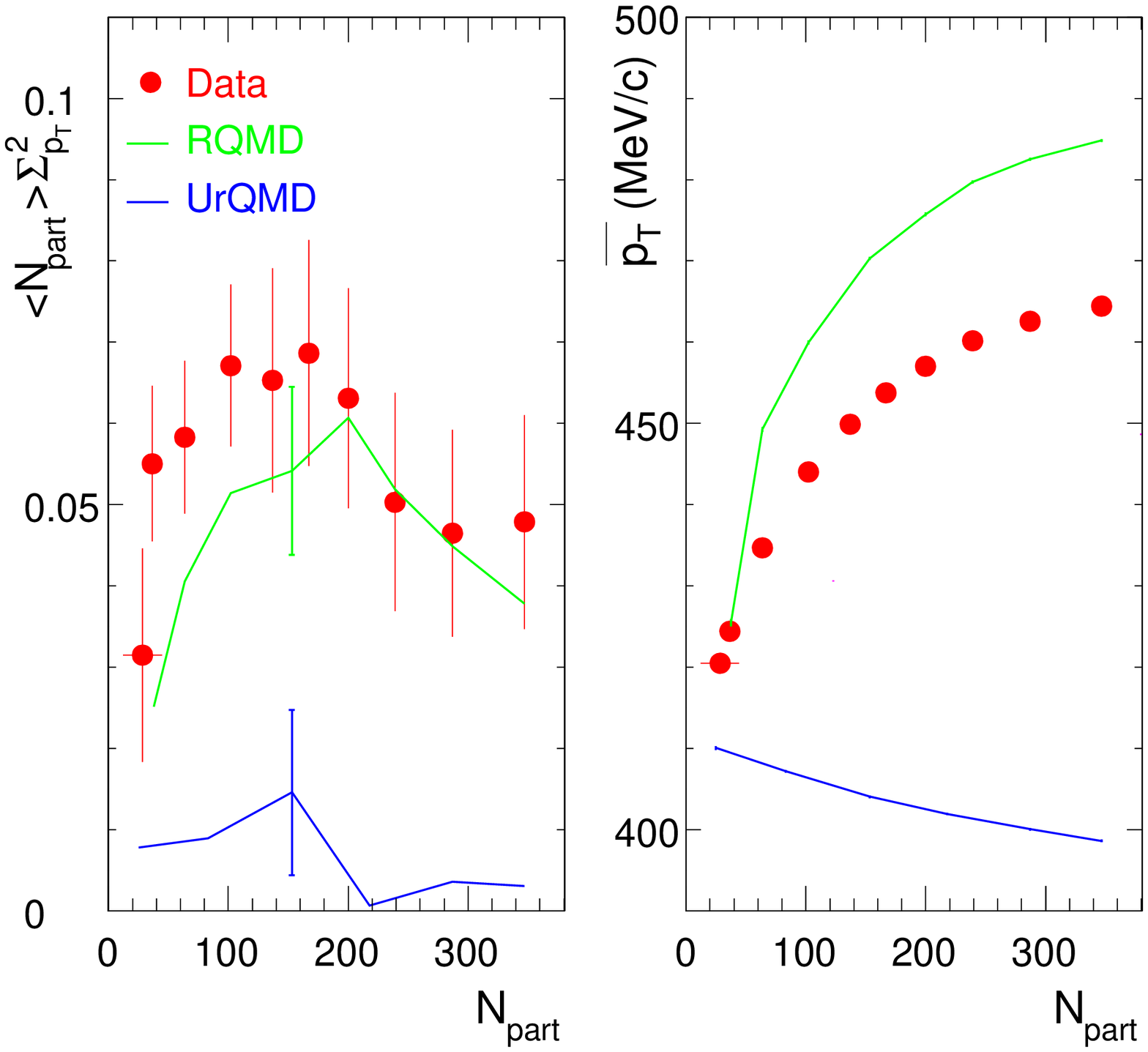}
\caption{Comparison of $M_{pt}$ fluctuations (left) and $\overline{p_t}$ (right)
at 158~$A$GeV/$c$ to {\sc rqmd} and {\sc urqmd}.}
\label{fig2}
\end{minipage}
\end{figure}

In Fig.\ref{fig1} (left) the results for $\Sigma_{pt}$ in 
minimum bias Pb-Au collisions at 40 and 158~$A$GeV/$c$ are shown as function
of $\langle N_{\rm part}\rangle$. 
Also shown are results from 
an analysis~\cite{ceresnpa} of central data sets at 40, 80, and 158~$A$GeV/$c$.
At all beam energies, a significant centrality dependence of $\Sigma_{pt}$
is observed. 
The magnitude of the fluctuations does not depend 
on beam energies.
In the most central events, $\Sigma_{pt}$ is about 1\% of $\overline{p_t}$,
consistent with previously reported results from SPS and 
RHIC~\cite{ceresnpa,star_westfall}.

As a reference, we employ an extrapolation from p-p data. At ISR, 
event-by-event fluctuations of $M_{pt}$ have been measured in 
p-p collisions at $\sqrt{s}$ between 30.8 and 63 GeV~\cite{braune}. 
For $\Sigma_{pt}^{pp}$,
a value of 12\% was found, with no significant dependence on
beam energy. 
Implying that particle production at SPS is approximately proportional to
$N_{\rm part}$, 
fluctuations may scale with the number $\langle N_{\rm part}\rangle$
of participating nucleons:
\begin{equation}
	\Sigma_{pt}^{AA} = \Sigma_{pt}^{pp}\left(\langle N_{\rm part}\rangle/2
	\right)^{-1/2}.
\end{equation}
As demonstrated in Fig.~\ref{fig1}, the data agree with this extrapolation
for very peripheral and central events. In contrast, a pronounced 
deviation is observed in semi-central events. 
In the right panel of Fig.~\ref{fig1}, the product 
$\langle N_{\rm part}\rangle \Sigma_{pt}^{2}$ 
is plotted as function of $\langle N_{\rm part}\rangle$. 
In this representation, the p-p extrapolation becomes a constant,
while the data exhibit a broad maximum around $N_{\rm part} = 120$. 
This observation is in qualitative agreement with previous findings
at SPS and RHIC~\cite{star_westfall,na49,star_ptfluc,phenix_ptfluc1}.

Non-statistical $M_{pt}$ fluctuations may be caused by momentum
space correlations which arise due to resonance decays. 
Moreover, the correlation strength is expected to increase
with the average transverse momentum of the decaying resonances.
We have therefore performed a comparison of our results at 158~$A$GeV/$c$ 
to the cascade models {\sc rqmd} and {\sc urqmd}.
The {\sc urqmd} model does not describe the observed increase of 
$\overline{p_t}$
with centrality, as demonstrated in Fig.\ref{fig2} (right).
Also, neither the magnitude nor the centrality dependence
of the $M_{pt}$ fluctuations can be reproduced by {\sc urqmd} (Fig.\ref{fig2},
left).
In contrast, {\sc rqmd} gives a rather good description of $\overline{p_t}$
and of the observed centrality dependence of $M_{pt}$ fluctuations.
This indicates that resonance decay kinematics combined with
the centrality dependent increase of $\overline{p_t}$ may give 
an important contribution to $M_{pt}$ fluctuations.
Further studies are needed, in particular regarding the failure of
{\sc urqmd}.

\subsection{Fluctuations of net charge}
\begin{figure}[thb]
\begin{minipage}[t]{89mm}
\includegraphics[width=\textwidth]{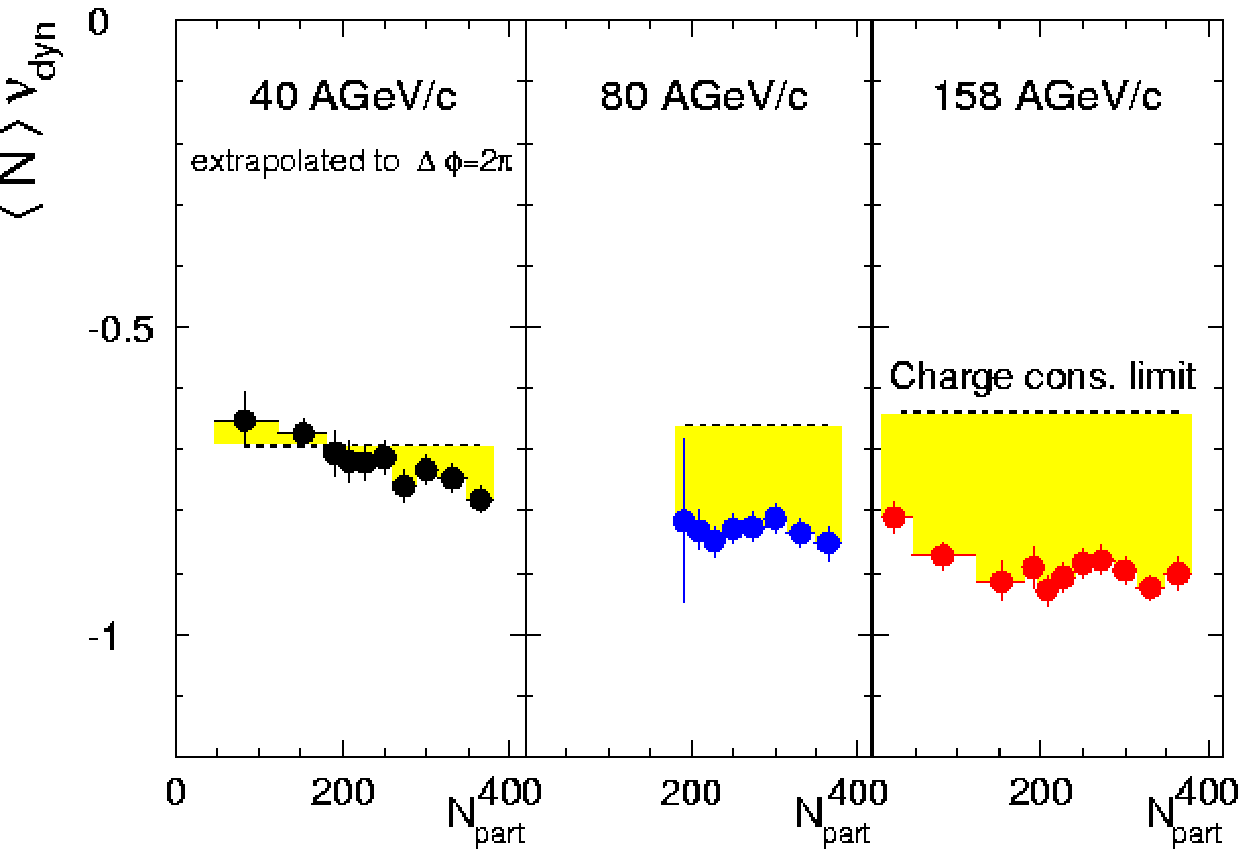}
\caption{The measure $\langle N\rangle \cdot \nu_{\rm dyn}$ as function
of centrality at different energies. Also shown is the GCC contribution.}
\label{fig3}
\end{minipage}
\hspace{\fill}
\begin{minipage}[t]{66mm}
\includegraphics[width=\textwidth]{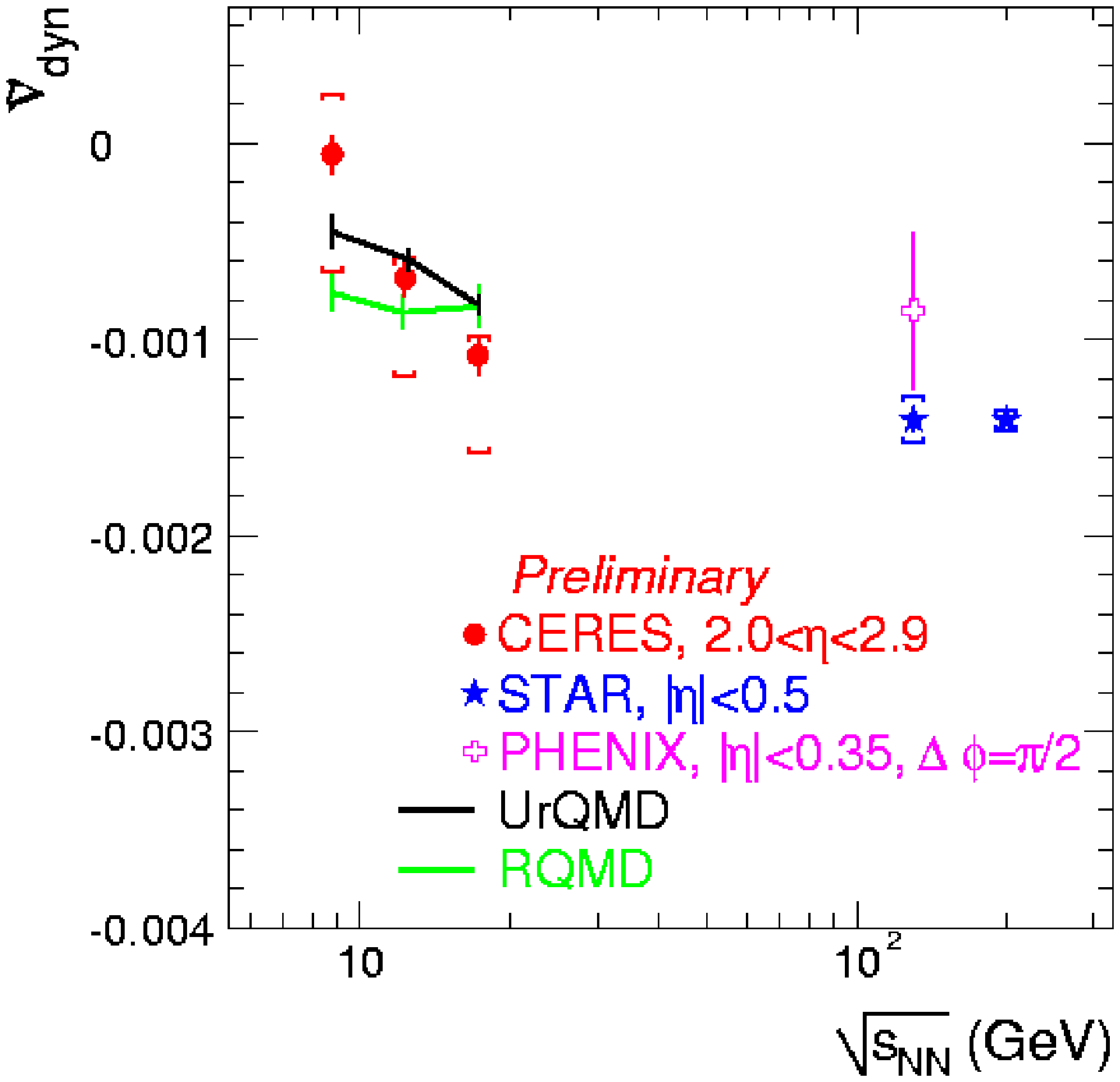}
\caption{Beam energy dependence of $\nu_{\rm dyn}$ after GCC correction.}
\label{fig4}
\end{minipage}
\end{figure}

As a measure for fluctuations of the net electric charge $Q=N_+ - N_-$
we evaluate the expression 
$\nu_{+-}=\langle \left(\frac{N_+}{\langle N_+\rangle}
-\frac{N_-}{\langle N_-\rangle}\right)^2\rangle$
and its statistical limit
$\nu_{\rm stat}=\left(\frac{1}{\langle N_+\rangle}+\frac{1}{\langle N_-\rangle}\right)$~\cite{gavin}.
Fig.\ref{fig3} shows the centrality dependence of 
$\nu_{\rm dyn}= \nu_{+-} - \nu_{\rm stat}$ multiplied by the mean multiplicity $\langle N\rangle$
of accepted particles at different beam energies. 
The results at 40~$A$GeV/$c$ are consistent 
with the expected contribution from global charge
conservation (GCC). 
With
increasing beam energy $\langle N\rangle\cdot\nu_{\rm dyn}$ decreases, 
indicating an increased
correlation between positive and negative particles.
At 158~$A$GeV/$c$, we note a decrease of $\langle N\rangle\nu_{\rm dyn}$
in peripheral collisions
which is connected to the narrowing of the charge balance function
in central collisions at SPS and at RHIC~\cite{star_westfall,gavin,marek}.

In Fig.\ref{fig4} is shown the beam energy dependence of $\nu_{\rm dyn}$
in central collisions, corrected for the contribution from GCC.
The result at top SPS energy is very similar to measurements at 
RHIC~\cite{star_westfall,star_charge,phen_charge}.
The good agreement of {\sc rqmd} and {\sc urqmd} with the data indicates
that the kinematics of resonance decays combined with the finite acceptance
of the experiment may be sufficient to
account for the observed charge fluctuations.
At all energies, the data exceed by far the expectation from an
equilibrated QGP of $\langle N\rangle \nu_{\rm dyn} \approx -3.5$.
It was, however, pointed out that in the framework of a constituent
quark coalescence model without gluons, the expected net charge fluctuations
as well as the evolution of the balance function are compatible with 
the data~\cite{bialas}.

\end{document}